\documentclass[twocolumn]{aastex631}

\usepackage{amsmath}
\usepackage{comment}

\emergencystretch=\maxdimen
\begin{document}

\newcommand{\scccc}[1]{\textcolor{red}{#1}}

\title{Cosmological Budget of Entropy from Merging Black Holes}
\correspondingauthor{Siyuan Chen}
\email{siyuan.chen@cfa.harvard.edu}
\author{Siyuan Chen}
\affiliation{Center for Astrophysics $\vert$ Harvard \& Smithsonian, 60 Garden Street, Cambridge, MA 02138, USA}
\affiliation{Department of Physics \& Astronomy, Vanderbilt University, 2301 Vanderbilt Place, Nashville, TN 37235, USA}

\author{Karan Jani}
\affiliation{Department of Physics \& Astronomy, Vanderbilt University, 2301 Vanderbilt Place, Nashville, TN 37235, USA}

\author{Thomas W. Kephart}
\affiliation{Department of Physics \& Astronomy, Vanderbilt University, 2301 Vanderbilt Place, Nashville, TN 37235, USA}

\begin{abstract}
Black holes contain more entropy than any other component of the observable universe. Gravitational-wave observations from LIGO and Virgo have shown evidence of a previously unknown black hole mass range, which provides new information to update the entropy budget. Increases in entropy due to binary black hole mergers, as implied in the second law of thermodynamics, should also be added to the budget. In this study, we update the cosmological entropy budget for black holes in the stellar to lite-intermediate-mass range $(5-300~M_\odot)$, originating from either supernovae or binary mergers, by utilizing a suite of population synthesis models and phenomenological fits derived from numerical relativity. We report three new insights: Firstly, the cumulative entropy from merging black holes surpasses the total entropy from cosmic microwave background photons around the onset of the Over-massive Black Hole Galaxy phase at $z\sim 12$, suggesting that mergers played a more significant role in shaping the thermodynamic state of the early universe than relic radiation. Secondly, if primordial black holes constitute a nonzero fraction of dark matter, their early binary mergers establish an ``entropy floor” in the Dark Ages and can dominate the cumulative merger-generated entropy history even for small abundances. Thirdly, by computing the cosmological density parameters, we highlight the thermodynamic asymmetry in black hole mergers, where the production of gravitational-wave energy is inefficient compared to the immense generation of Bekenstein-Hawking entropy.

\end{abstract}


\section{Introduction} \label{sec:intro}

Most gravitational wave (GW) events reported by the LIGO-Virgo-KAGRA (LVK) collaboration are associated with binary black hole (BBH) mergers \citep{LIGOScientific:2016vlm, yunes2024gravitationalwavetestsgeneralrelativity}.
According to the second law of thermodynamics, irreversible processes, such as black hole mergers \citep{Kephart_2003}, would increase the entropy of the universe \citep{PhysRevD.7.2333}. As the universe evolves toward greater disorder through rising entropy, this thermodynamic progression is closely connected to the arrow of time. Therefore, entropy offers a novel lens for probing the evolution of the universe. 

Understanding this thermodynamic evolution requires placing it in the context of the universe’s expansion history. As outlined by \cite{Frieman_2008}, the universe has evolved through three distinct eras, each dominated by a different constituent. At $z \gtrsim 3000$, shortly after the Big Bang, the universe was in the radiation-dominated era.  This was followed by the matter-dominated era, spanning approximately $0.5 \lesssim z \lesssim 3000$, during which the gravitational influence of matter shaped the large-scale structure of the universe. The current epoch, defined by $z \lesssim 0.5$, is governed by dark energy, which drives the accelerated expansion of the universe. These cosmic epochs are visually marked in our plots, where we compare the entropy evolution across the matter- and dark energy-dominated eras to extract new insights into the thermodynamic behavior of the universe and its large-scale structure formation across cosmic time.

Past studies including \cite{Frampton_2008, Berman_2009, Frampton_2009, frampton2009entropyintermediatemassblackholes,  Egan_2010, Frampton_2011} have estimated the entropy budget associated with various cosmic components. However, numerous observational advances over the past decade have significantly refined our understanding of black hole populations and their contributions to the entropy budget of the universe. In particular, \cite{Egan_2010} categorized heavier stellar-mass black holes in the mass range $42 - 140~M_\odot$ as \textit{tentative components} due to their overlap with the so-called \textit{upper mass gap} - more commonly referred to in the literature as the Pair-Instability Supernovae (PISN) mass gap. Stellar evolution and nucleosynthesis theories predict the existence of a mass gap, spanning approximately $ 45 - 130~M_\odot$ \footnote{The precise bounds of the PISN mass gap remain uncertain.}, in which black holes cannot form \textit{directly}. Progenitors entering this regime undergo Pulsational Pair-Instability Supernovae (PPISN), shedding significant mass before collapse and potentially forming a pile-up of black holes just below the gap. For more massive cores, the instability leads to a full PISN explosion that completely disrupts the star, leaving no compact remnant behind \citep{Woosley_2017, Farmer_2019, Fishbach_2020, Marchant_2020, Woosley_2021, Edelman_2022}. However, observations from the LVK collaboration challenge this exclusion picture \citep{Di_Carlo_2020, O_Brien_2021, Costa_2022, winch2024predictingheaviestblackholes}. The most prominent example, GW190521 \citep{LIGOScientific:2020iuh}, involves two inspiral black holes ($m_1 = 85^{+21}_{-14}~M_{\odot}$ and $m_2 = 66^{+17}_{-18}~M_{\odot}$) that reside squarely within the PISN gap, producing an intermediate-mass black hole (IMBH) remnant of $m_\mathrm{f} = 142^{+28}_{-16}~M_{\odot}$ \citep{GW190521astro, Greene_2020, Ruiz-Rocha2025yno}. Furthermore, the most massive event reported to date, GW231123 \citep{2025ApJ...993L..25A}, features a primary component ($m_1 = 137^{+23}_{-18}~M_\odot$) and a remnant ($m_\mathrm{f} = 222^{+28}_{-42}~M_\odot$) that sit at or beyond the upper edge of the predicted gap.

Recent studies, such as \cite{profumo2024newcensusuniversesentropy}, have provided updated assessments of entropy contributions from both known and previously overlooked components including IMBHs. Building upon the foundational census of \cite{Egan_2010}, this study integrates the latest findings from GW astronomy and cosmological observations. In Sec. \ref{subsec:merging_bud}, we present a revised estimate of the total cosmological entropy, offering a comprehensive and up-to-date inventory that includes these newly constrained populations. Furthermore, in Sec. \ref{sec:cumulation+rate}, we analyze these results in the context of the Cosmic Microwave Background (CMB) and Primordial Black Holes (PBHs), providing new insights into the thermodynamic evolution of the universe.

\section{Methods} \label{sec:method}


\subsection{Entropy for Stellar Black Holes}
\label{subsec:bh_imf}
Based on the Hawking Area Theorem \citep{PhysRevLett.26.1344, Sarkar:2017qln, Isi_2019, Isi_2021, kontou2023generalizationhawkingblackhole, li2023testsnohairtheorembinary}, the entropy of a black hole correlates to the surface area of its event horizon. By expressing the Bekenstein-Hawking Formula \citep{PhysRevD.7.2333} in terms of mass (m) and spin (a), we can compute the entropy of an individual black hole as: 
\begin{equation}
\label{eq:entropy}
\centering
    S = \frac{kA}{4l_p^2} = \frac{kAc^2}{4G\hbar^2} = \frac{2\pi k G}{\hbar c} m^2 (1+\sqrt{1-a^2})
\end{equation}


\begin{figure*}[t!]
\centering
\includegraphics[trim = 0 10 0 0, clip, width=0.98\textwidth]{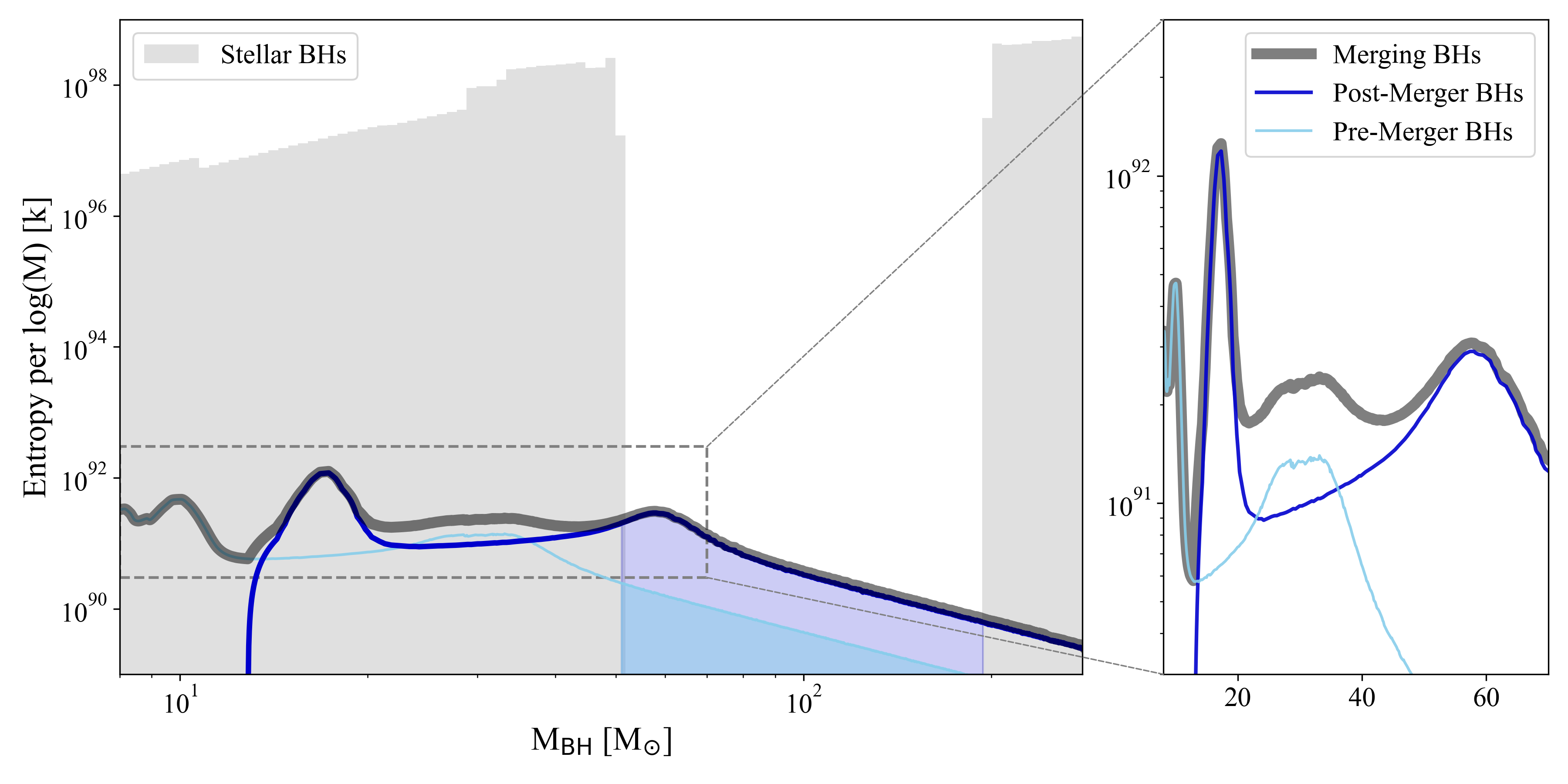}
\caption{\textbf{Cosmological Binary Black Hole Entropy Inventory}: the total accumulated entropy (Static Budget) from different black hole populations as a function of black hole mass. The stellar–origin black holes (grey histogram) include black holes formed via supernova and direct collapse. The population of merging black holes is decomposed into pre-merger components (light blue) and post-merger remnants (dark blue), both contributing in the range $5-200~M_\odot$. The total entropy budget from merging binaries is shown as a thick gray band. The right panel zooms in on the entropy profile of merging black holes in the range $8-70~M_\odot$ to highlight the detailed structure.}
\label{fig:imf}
\end{figure*}

In this study, we categorize stellar-origin black holes (remnants of stellar collapse) into three specific regimes. We define standard \textit{stellar BHs} as having $5 -45~M_\odot$. Objects residing within the PISN mass gap, $45-130~M_\odot$, are referred to as \textit{PISN BHs}. Very massive stars (VMS, $\gtrsim 260~M_\odot$) may instead collapse directly, bypassing the PISN mechanism and producing remnants in the range of $130-300~M_\odot$, which we denote as \textit{lite-IMBHs} \citep{Woosley_2017, Ezquiaga_2021, Mehta_2022, franciolini2024farsideobservingblack, Ruiz-Rocha2025yno, Chatterjee_2025}. More broadly, we adopt $10^2 - 10^7~M_\odot$ as the \textit{IMBHs} range. Finally, black holes exceeding $10^7~M_\odot$ are classified as supermassive black holes (\textit{SMBHs}). Although rare in number, SMBHs are expected to be the dominant contributors to the total cosmic entropy budget. All entropies are expressed in units of $k$ and compared on a common comoving basis.

To compute the total number density of massive stars that can collapse into black holes, we adopt the Salpeter Initial Mass Function \citep[IMF;][]{1955ApJ...121..161S} parameterized by a power-law index $\alpha = 2.35^{+0.35}_{-0.65}$:

\begin{equation}
\xi(m_\star)~dm = \xi_0 \left( \frac{m_\star}{M_\odot} \right)^{-\alpha} dm,
\end{equation}
where $\xi_0$ is the normalization constant. We calibrate $\xi_0$ by requiring that the integral of the mass-weighted IMF over the full stellar mass range (assumed to be $0.1 - 350\,M_\odot$) reproduces the observed present-day cosmic stellar mass density, $\Omega_\star \simeq 0.0027\pm0.0005$ \citep{Fukugita_2004, Madau_2014}:

\begin{equation}
  \rho_\star = \int^{350~M_\odot}_{0.1~M_\odot} m_\star\,\xi(m_\star)\,dm_\star = \Omega_\star\,\rho_{\rm c,0}
\end{equation}
where $\rho_{\rm c,0}$ is the critical density of the universe at $z=0$, assuming a flat $\Lambda$CDM cosmology \citep{2020A&A...641A...6P}.

With $\xi_0$ determined, the number density of black hole progenitors within the specific mass interval $[m_{\rm min}, m_{\rm max}] = [8, 350]\,M_\odot$ is given by:
\begin{equation}
    n_\star = \int^{m_\mathrm{max}}_{m_\mathrm{min}} \xi(m_\star)~dm_\star
    \label{eq:numberimfstars}
\end{equation}

For each progenitor mass $m_\star$, we determine the corresponding remnant mass, $m_{\rm rem}$, using the SEVN stellar evolution code \citep{Spera_2017}. We assume a low-metallicity environment ($Z=2\times10^{-4}$) to account for the formation of massive black holes and include the effects of PPISN and PISN.

We note that the mapping $m_\star \mapsto m_{\rm rem}$ is non-bijective due to mass-loss pile-ups (PPISN) and complete disruptions (PISN). Consequently, a simple Jacobian transformation of the IMF is invalid \citep{Jani_2020}. Instead, we compute the remnant mass function, $\xi_{\rm BH}(m_{\rm rem})$, by enforcing particle number conservation: we sum the IMF contributions from all disjoint progenitor intervals that map into a given remnant mass bin.

To determine the total number density of black holes and their cumulative contribution to the cosmic entropy budget, we integrate the derived remnant mass function, $\xi_{\rm BH}(m_{\rm rem})$, over the remnant mass range. For the entropy calculation, we require a spin assignment for each black hole. We assume a uniform distribution for the dimensionless spin parameter, $a \sim \mathcal{U}[0,1]$, and compute the entropy for each remnant mass using Eq. \ref{eq:entropy}. The resulting distributions of remnant count and total entropy are illustrated by grey histograms in Fig. \ref{fig:imf}.

\subsection{Entropy for Merging Binary Black Holes}

For merging BBHs, we can get 4 different types of entropies: entropy of the primary black hole ($S_1$), entropy of the secondary black hole ($S_2$), entropy of the remnant black hole ($S_\mathrm{f}$), and the entropy produced during the merger ($\Delta S$), which can be calculated by:

\begin{equation}
\label{eq:delta-entropy}
\centering
    \Delta S = S_\mathrm{f} - (S_1 + S_2)
\end{equation}

\begin{figure}[hbt!]
\centering
\includegraphics[width=0.49\textwidth]{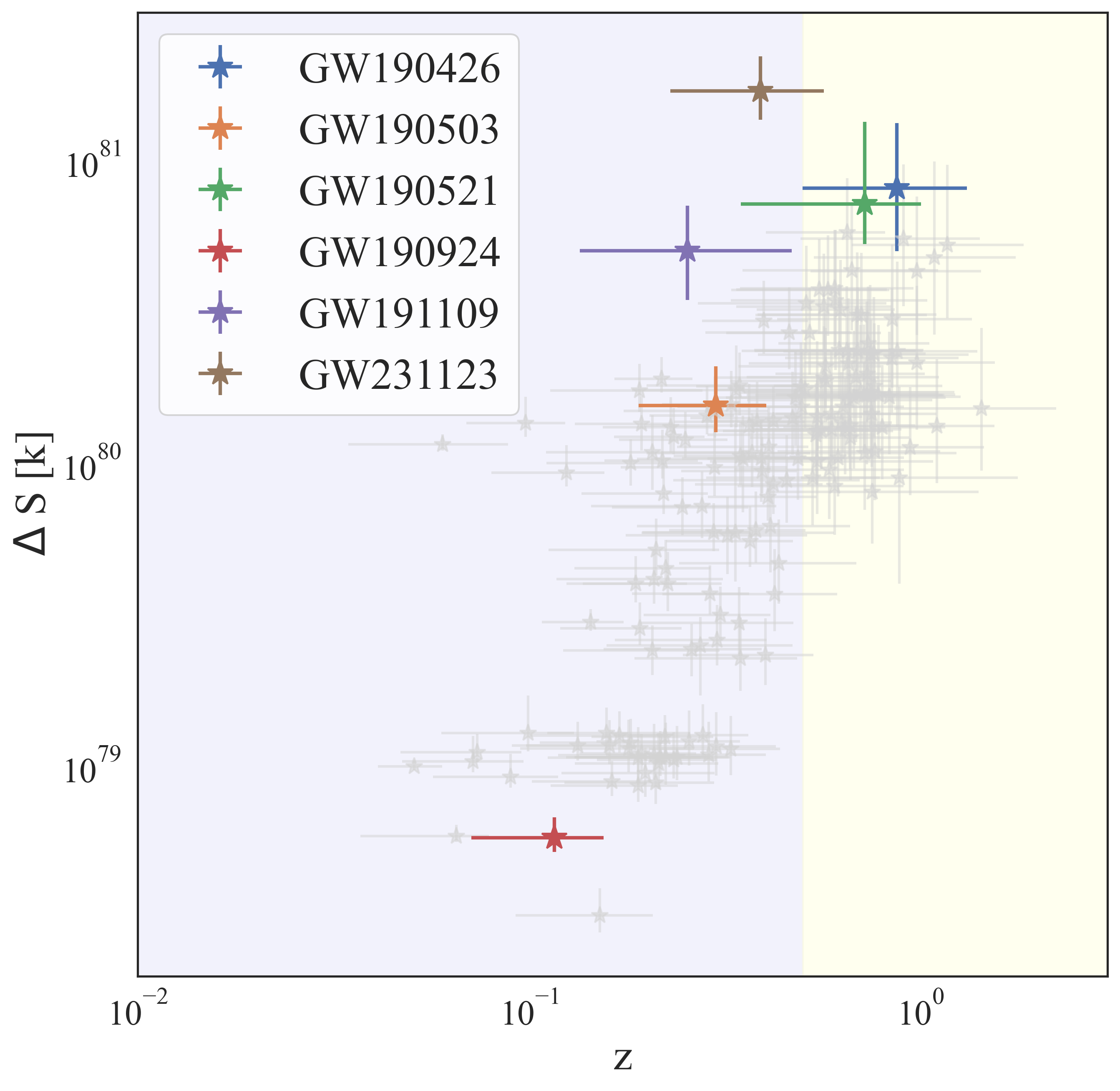}
\caption{\textbf{Entropy Production in Individual GW Events}: grey bars show the correlation between the redshift of the binary and the change in entropy during BBH mergers for 168 BBH events reported by LVK. The bar represents $z$ and $\Delta~S$ with 90\% confidence interval. Certain GW events are highlighted in color. The purple-shaded region corresponds to the dark energy–dominated era ($z \le 0.5$), while the yellow-shaded region represents the matter-dominated era ($z > 0.5$).}
\label{fig:lvk}
\end{figure}

To compute the entropy change for individual BBHs reported by LVK, we utilize the parameter estimation results made available in the GW Open Science Center \citep{2021SoftX..1300658A, 2023ApJS..267...29A}. We adopt posterior samples obtained with the IMRPhenomXPHM waveform model \citep{Khan_2016, Khan_2019, Pratten_2020, Pratten_2021, dambrosio2024testinggravitationalwaveformsgeneral}. These samples include the mass and spin associated with the corresponding remnant black hole, which has been computed using phenomenological fits to numerical relativity (NR) results \citep{Santamar_a_2010, Hofmann_2016, Healy_2017, Jim_nez_Forteza_2017,  Estell_s_2022}. For each posterior sample, we evaluate Eq. \ref{eq:entropy} for the primary, secondary, and remnant black holes to obtain ($S_1, S_2, S_\mathrm{f}$) and then compute the entropy change $\Delta S$ with Eq. \ref{eq:delta-entropy}. The resulting event-level $\Delta S$ measurements are summarized in Fig. \ref{fig:lvk}.

To extend the analysis to the population level, we adopt the Power Law + Peak (PP) model \citep{Talbot_2018}, which demonstrates the highest Bayes factor in characterizing the Gravitational-Wave Transient Catalog dataset of the Third / Fourth Observing Run (GWTC-3 / 4). The detailed parameter distributions are described in \cite{gwtc-3-pop, gwtc4pop}. To propagate population-level uncertainties into our entropy estimates, we evaluate results using representative lower and upper bounds of the BBH merger rate inferred by LVK (as reported in the catalog analyses). This brackets the range of plausible merger-driven entropy production within the model and catalog uncertainties.

Given the hyperparameterized distributions of the PP model, we generate a Monte Carlo sample of $N_\mathrm{BBH}$ synthetic binaries by drawing inspiral parameters $\Lambda^\mathrm{ins}_i: \{ m_{1,2},~a_{1,2},~\theta_{1,2} \}$ with bounds determined by the adopted population model. For each draw $\Lambda^\mathrm{ins}_i$, we compute the remnant parameters $\Lambda^\mathrm{rem}_i: \{ m_\mathrm{f},~a_\mathrm{f} \}$ using the NR surrogate fit \textsc{NRSur7dq4Remnant} \citep{Blackman_2017, Boschini_2023} as implemented in the \textit{SurfinBH} package \citep{soft09007V, Varma_2019, zertuche2024highprecisionringdownsurrogatemodel}. We then evaluate Eq. \ref{eq:entropy} to obtain ($S_1, S_2, S_\mathrm{f}$) and Eq. \ref{eq:delta-entropy} to compute the corresponding entropy change $\Delta S$.

\label{subsec:mergerrate}
Observational constraints from GWTC-3/4 population analyses parameterize the BBH merger-rate density as a redshift-dependent power law $\mathcal{R}_\mathrm{BBH}(z) \propto (1+z)^\kappa$ \citep{gwtc-3-pop, gwtc4pop}. Owing to the sensitivity of current ground-based detectors, this evolution is directly constrained primarily in the local universe, roughly over $z \in [0.2, 1.5]$. For $z \ge 1.5$, we utilize $\mathcal{R}_\mathrm{BBH}(z)$ reported in \cite{Mapelli_2018} derived from the Illustris simulation. In this work, we discretize redshift into bins uniform in lookback time with $\Delta t_\mathrm{lb}= 10^7~yr$. 

The merger rate may be written as $\mathcal{R}(z) = dN_\mathrm{BBH}/(dV_c~dt)$ \citep{Mandel_2022, gwtc-3-pop}. The expected number of BBH mergers in a lookback-time bin centered at redshift $z$ is then
\begin{equation}
    \centering
    \label{eq:growth}
    N_\mathrm{BBH}(z) = \mathcal{R}(z) \times \Delta V_c(z) \times \Delta t_\mathrm{lb}(z)
\end{equation}
where $\Delta V_c$ is the differential comoving volume element associated with the bin and $\Delta t_\mathrm{lb}$ is the corresponding lookback-time width. For each bin, we draw $N_\mathrm{BBH}(z)$ sets of inspiral parameters $\Lambda^\mathrm{ins}_i$ from the adopted population model and compute the corresponding entropy changes $\Delta S_i$. The total merger-generated entropy in that bin is therefore $\sum_{i=1}^{N_\mathrm{BBH}} \Delta S_i$. \footnote{For computational efficiency, we only sample and compute the total entropy change of $10^6$ mergers and scale with $N_\mathrm{BBH}$.}

We also consider the corresponding entropy density, defined as
\begin{equation}
    \centering
    s = \frac{S}{V_\mathrm{obs}}
    \label{eq:density}
\end{equation}
where $V_\mathrm{obs}$ is the volume of the observable universe. We adopt $3.52^{+0.11}_{-0.11} \times 10^{80}~m^3$ \citep{profumo2024newcensusuniversesentropy}. 

\begin{table*}[t!]
\caption{Updated entropy budget table of different cosmological components in the observable universe, with comparisons to previous work. The IMBH category remains poorly constrained observationally, leading to large uncertainties in entropy estimates. All entropy densities computed by dividing by $V_\mathrm{obs}$. Reference: 
[1] \cite{Frautschi1982}, 
[2] \cite{Frampton_2008},
[3] \cite{frampton2009entropyintermediatemassblackholes},
[4] \cite{Frampton_2009},
[5] \cite{Egan_2010},
[6] \cite{profumo2024newcensusuniversesentropy}
}

\label{tab:budget}

\resizebox{\textwidth}{!}{%
\begin{tabular}{llll}
\hline\hline
\textbf{Components} & \textbf{Entropy Density $s$ [$k~\mathrm{m}^{-3}$]} & \textbf{Entropy $S$ [$k$]} \ \ \ \ & \textbf{Entropy $S$ [$k$] } \\
 &   & \textbf{(This Work)} & \textbf{(Previous Work)} \\
\hline 
Stellar BHs ($5{-}45~M_\odot$) & $1.8 \times 10^{18^{+0.8}_{-2.1}}$ & $2.8 \times 10^{98^{+0.8}_{-2.1}}$ & $10^{96}$ [6], $10^{97}$ [4][5], $10^{98}$ [1] \\

PISN BHs ($45{-}130~M_\odot$) & $1.2^{+1.3}_{-0.5} \times 10^{12}$ & $1.9^{+2.1}_{-0.7} \times 10^{93}$ & $10^{99}$ [5]\\

Lite-IMBHs ($130{-}300~M_\odot$) & $5.1 \times 10^{18^{+0.9}_{-1.6}}$ & $1.8 \times 10^{99^{+0.9}_{-1.6}}$ & - \\

IMBHs ($10^2{-}10^7~M_\odot$) & $1.4 \times 10^{17^{+8.3}_{-1.2}}$ [3][6] & - &  $10^{97}$ [6], $10^{105}$ [3] \\
SMBHs ($10^7{-}10^9~M_\odot$) & $2.4^{+3.0}_{-2.2} \times 10^{21}$ [6] & - & $10^{101}$ [6], $10^{102}$ [4], $10^{103}$ [2], $10^{104}$ [5]\\

\hline
BBH mergers $\Delta S$ ($5{-}\sim 500~M_\odot$) & $7.1^{+9.2}_{-3.5} \times 10^{12}$ & $1.1^{+1.5}_{-0.6} \times 10^{93}$ & - \\
GWTC Pre-merger $S$ ($5{-}300~M_\odot$) & $1.2^{+1.5}_{-0.6} \times 10^{13}$ & $1.9^{+2.5}_{-0.9} \times 10^{93}$ & - \\
GWTC Post-merger BHs ($8{-}\sim 500~M_\odot$) & $1.9^{+2.5}_{-0.9} \times 10^{13}$ & $3.1^{+4.0}_{-1.5} \times 10^{93}$ & - \\

\hline
Photons & $1.5 \times 10^{9}$ [6] & - & $10^{88}$ [4], $10^{89}$ [5][6]\\
Relic Neutrinos & $1.4 \times 10^{9}$ [6] & - & $10^{88}$ [4], $10^{89}$ [5][6]\\
WIMP Dark Matter & $8.9^{+64}_{-6.0} \times 10^{7}$ [6] & - & $10^{88}$[5], $10^{87}{-}10^{89}$[6]\\
Relic Gravitons & $1.7 \times 10^{{7}^{+3.5}_{-2.5}}$ [5][6] & - & $10^{87}$ [5], $10^{87}{-}10^{91}$[6]\\
ISM and IGM & $7.5^{+1.1}_{-7.0} \times 10^1$ [5][6] & - & $10^{81}$[5], $10^{80}{-}10^{81}$[6]\\
Stars & $2.6\times10^{{-1}^{+4.5}_{-3.7}}$ [5][6] & - & $10^{79}$ [4],$10^{80}$ [5], $10^{74}{-}10^{81}$[6]\\
\hline\hline
\end{tabular}%
}
\end{table*}


\subsection{Entropy Budget for Primordial Black Holes}
To contextualize the entropy contribution of stellar-origin black holes, we consider the potential contribution from PBHs. PBHs are hypothetical compact objects formed in the radiation-dominated era ($z \gtrsim 3000$), which may constitute a fraction $f_\mathrm{PBH} \equiv \Omega_\mathrm{PBH} / \Omega_\mathrm{DM}$ of the total dark matter density, where $\Omega_\mathrm{DM} \approx 0.26$ \citep{2020A&A...641A...6P}. While current constraints from LVK merger rates suggest an upper limit of $f_\mathrm{PBH} \lesssim 10^{-3}$ for stellar-mass PBHs \citep{Sasaki_2016, H_tsi_2021}, even a trace population of these objects can significantly impact the cosmic entropy budget, particularly at high redshifts.

We adopt a log-normal mass distribution for the PBH population, which is a standard prediction for formation from smooth, symmetric peaks in the primordial power spectrum:
\begin{equation}
    \psi(M) = \frac{1}{\sqrt{2\pi}\sigma M} \exp\left[ -\frac{\ln^2(M/M_c)}{2\sigma^2} \right]
\end{equation}
where we assume a characteristic mass $M_c = 30~M_\odot$ and a width $\sigma = 0.5$. This mass range is chosen to be consistent with the component masses of BBHs observed by LVK \citep{KAGRA:2021duu, gwtc-3-pop, gwtc4pop}. Unlike stellar-origin black holes, which can acquire significant spin from their progenitors, PBHs are expected to form with negligible natal spin due to the efficiency of radiation drag during the early universe \citep{Raidal_2019}. We therefore assume a dimensionless spin parameter $a = 0$ for the entire PBH population.

The merger history of PBHs differs fundamentally from that of stellar binaries. While stellar mergers follow the star formation rate (peaking at $z \sim 2$), the PBH merger rate is driven by the dynamics of binaries formed in the early universe. We adopt the merger rate density derived by \citet{Sasaki_2016}:
\begin{equation}
    \mathcal{R}(t) = \mathcal{R}_0 \left( \frac{t}{t_0} \right)^{-34/37},
\end{equation}
where $t$ is the cosmic time and $t_0$ is the current age of the universe. The power-law index $-34/37 \approx -0.92$ implies that the rate increases monotonically with redshift. The normalization factor $\mathcal{R}_0$ depends non-linearly on the PBH abundance, reflecting the probability of binary formation:
\begin{equation}
    \mathcal{R}_0 \approx 1.6 \times 10^6 \left( \frac{f_\mathrm{PBH}}{10^{-3}} \right)^{53/37} ~\mathrm{Gpc}^{-3}\,\mathrm{yr}^{-1}.
\end{equation}
This scaling ($f_\mathrm{PBH}^{1.43}$) indicates that even a small decrease in $f_\mathrm{PBH}$ results in a sharp drop in the merger rate, providing a natural mechanism to satisfy observational constraints while still allowing for a dominant entropy contribution in the dark ages.

\subsection{Cosmological Density Parameters}
While $\Omega$ quantifies the associated mass–energy density relative to the critical density, the Bekenstein–Hawking entropy weights that inventory nonlinearly through the area law. Therefore, astrophysical populations, especially compact objects, can contribute negligibly to $\Omega$ yet dominate the entropy budget.

Alongside entropy, we report the cosmological energy budget of black holes and their merger products using the dimensionless density parameter, $\Omega_i$. For a cosmic component $i$ with mass-energy density $\rho_i$, the density parameter is defined as:
\begin{equation} 
\Omega_i = \frac{\rho_i}{\rho_{c,0}}, \quad \text{where} \quad \rho_{c,0} = \frac{3H_0^2}{8\pi G} 
\end{equation}
is the critical density of the universe at the present epoch ($z=0$). We derive three specific parameters to characterize the black hole population: the initial mass density ($\Omega_\mathrm{BH}$), the gravitational-wave energy density ($\Omega_\mathrm{GW}$), and the post-merger remnant density ($\Omega_\mathrm{BH,rem}$).

The parameter $\Omega_\mathrm{BH}$ represents the total mass stored in stellar-origin black holes prior to any binary mergers. It serves as the initial ``mass inventory" of the compact object sector. We calculate this by integrating the remnant mass function, $\xi_\mathrm{BH}(m)$, derived in Section \ref{subsec:bh_imf}, over the full mass range of the population:
\begin{equation} 
\rho_\mathrm{BH} = \int_{m_\mathrm{min}}^{m_\mathrm{max}} m~\xi_\mathrm{BH}(m)~dm \end{equation}

During the coalescence of BBHs, a fraction of the total rest-mass energy is radiated as GWs by global energy conservation: $M_\mathrm{tot} c^2 = M_\mathrm{rem} c^2 + E_\mathrm{GW}$. To derive the cosmic energy density of this stochastic background, $\rho_\mathrm{GW}$, we integrate the energy released by the merger rate density $\mathcal{R}(z)$ over cosmic history. We account for the expansion of the universe, which dilutes the energy density of radiation emitted at redshift $z$ by a factor of $(1+z)$ as it propagates to the present day. The total energy density is thus:
\begin{align}
    \rho_\mathrm{GW} &= \frac{1}{c^2} \int^{\infty}_0 \frac{\mathcal{R}(z) \langle E_\mathrm{GW} \rangle}{1+z} \frac{dt}{dz}dz\\
    &= \int^{\infty}_0 \frac{\mathcal{R}(z) \langle \epsilon_\mathrm{GW} M_\mathrm{tot}\rangle}{1+z} \frac{dt}{dz}dz
\end{align}

We compute the radiative efficiency $\epsilon_\mathrm{GW}$ using the phenomenological fit \textsc{NRSur7dq4Remnant}, following the methodology outlined in Sec. \ref{subsec:mergerrate}. Finally, $\Omega_\mathrm{BH,rem}$ quantifies the mass density retained in remnant black holes after merger, with the progenitor mass–energy budget partitioned into $\Omega_\mathrm{BH,rem}$ and $\Omega_\mathrm{GW}$.


\section{Results} 
\subsection{Entropy Budget for Stellar Black Holes}
In Fig. \ref{fig:imf}, we presented the entropy budget of black holes distributed across logarithmic mass intervals. Throughout the discussion, the ``entropy budget” refers to the sum of \textbf{Bekenstein–Hawking entropies} of black holes. We considered two complementary estimates: an IMF-based, cosmological inventory for stellar remnants (Sec. \ref{subsec:bh_imf}), and a LVK population-model-based budget for BBH components and remnants (Sec. \ref{subsec:mergerrate}). For stellar black holes with masses below $45~M_\odot$, the number density decreases with increasing mass, yet the total entropy continues to rise due to the quadratic dependence of black hole entropy on mass. The cumulative entropy budget of this population is approximately $2.8 \times 10^{98}~k$. In comparison, both number density and entropy decrease with mass for lite-IMBHs. The total entropy associated with this population is $1.8 \times 10^{99}~k$, slightly exceeding that of the stellar-mass black holes due to their significantly larger individual entropies despite their lower abundance. 

Theoretical models of PISN predict a mass gap that strongly suppresses the formation of compact remnants between approximately $ 45 - 130~M_\odot$. However, GW observations reveal the existence of compact objects within this nominally disfavored region. Specifically, we identify three distinct populations of black holes that can populate the mass gap:

\begin{enumerate}
    \item Pre-merger primary black holes ($S_1$): according to the mass distribution inferred from GWTC-4, $1.3^{+0.2}_{-0.1}~\%$ of primary black holes fall within the mass range $ 45 - 130~M_\odot$. The associated entropy budget for these black holes is $2.0^{+2.7}_{-1.0} \times 10^{92}~k$.
    \item Pre-merger secondary black holes ($S_2$): slightly fewer secondary black holes reside in the PISN mass gap, so their entropy budget is comparatively lower, estimated at $9.3^{+14}_{-5.1} \times 10^{91}~k$.
    \item Post-merger black holes ($S_\mathrm{f}$): remnant black holes are significantly more likely to form within the PISN mass gap ($> 50\%$). As a result, they contribute the largest fraction of entropy among PISN black holes, with a total entropy of $1.6^{+1.7}_{-0.7} \times 10^{93}~k$. 
\end{enumerate}

Despite their low number density ($\sim 1.3\%$), PISN black holes contribute $15^{+0.7}_{-0.8}~\%$ of the total entropy of the pre-merger black hole population. In contrast, for remnant black holes, PISN black holes account for a dominant $51^{+6.0}_{-4.1}~\%$ of the total entropy budget. As shown in Fig. \ref{fig:imf}, the majority of entropy budget in the mass gap comes from post-merger black holes (blue).

The entropy distribution of the black hole population inherits the characteristic features of the PP model. In the right panel of Fig. \ref{fig:imf}, we resolve the fine structure of these components. The primary component black holes ($m_1$) establish the foundational peaks at $10~M_\odot$ and $35~M_\odot$. The contribution from $m_2$ manifests differently across the mass spectrum: at the low-mass end, they form a distinct subsidiary peak at $\sim 8~M_\odot$; conversely, near the pair-instability pile-up, the secondary population broadens the $35~M_\odot$ peak into a flattened plateau structure. Finally, the post-merger remnant population generates two prominent peaks at $\sim20~M_\odot$ and $\sim 70~M_\odot$.

We observe that the entropy budget of PISN black holes is approximately $10^6$ times smaller than that of stellar-mass black holes and light-IMBHs formed from VMSs. The detailed values are shown in Table \ref{tab:budget}. This significant difference is consistent with the expectation that the formation channels for black holes within the PISN mass gap are comparatively rare and require specialized pathways \citep{Costa_2022, Karathanasis_2023, chen2024distinguishingdemographicscompactbinaries}. 

In our calculation, we find that the entropy budget of light stellar-mass black holes ($\leq 15~M_\odot$) formed via core collapse supernova and fallback is $\sim 6.5 \times 10^{97}~k$, which lies within the uncertainty range $5.9 \times 10^{{97}^{+0.6}_{-1.2}}~k$ reported by \cite{Egan_2010}. Similarly, the entropy budget for IMBHs formed from VMSs is also consistent with the results presented in the same study.

\subsection{Entropy Budget for Merging Black Holes}
\label{subsec:merging_bud}
In this section, we focus on the entropy increase $\Delta S$ associated with BBH mergers. We present the entropy increase inferred for 168 LVK events broadly classified as BBH systems in Fig. \ref{fig:lvk}. The data show a weak correlation between redshift and entropy generation, with a R-squared $R^2\approx0.15$. Previous studies suggest that current observations increasingly favor higher-mass systems at higher redshifts \citep{heinzel2023probingcorrelationsbinaryblack, heinzel2024nonparametricanalysiscorrelationsbinary, Rinaldi_2024}, primarily due to selection effects.  Since black hole entropy scales quadratically with mass, this observational bias would naturally induce an increase in entropy with redshift, offering a plausible explanation for the observed weak trend. However, detecting compact object coalescences at redshifts $z \geq 2$ remains a significant challenge for current ground-based GW detectors. Within the accessible range of $z \in [0,2]$, no statistically significant correlation between redshift and entropy change $\Delta S$ is evident. Given this lack of strong evolutionary evidence, we assume that the underlying black hole mass function is redshift-independent. Therefore, we adopt the PP model uniformly across the entire redshift range of our analysis ($z \in [0,20]$).


\begin{figure}[t!]
\centering
\includegraphics[width=0.48\textwidth]{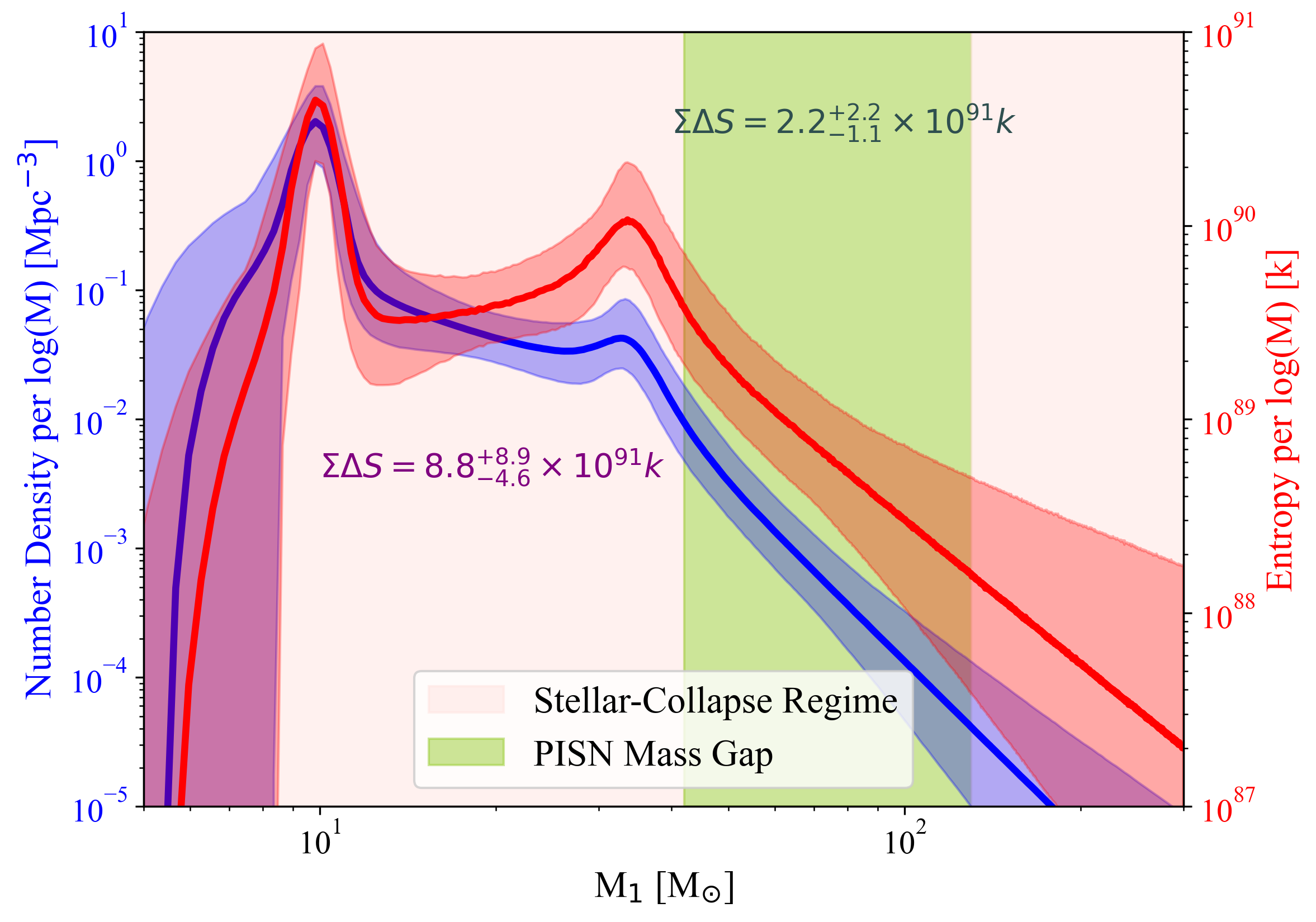}
\caption{\textbf{Dynamic Entropy Production}: The blue curve (left axis) shows the GWTC-4 inferred number density of BBH mergers per Mpc$^{-3}$ per logarithmic mass interval. The red curve (right axis) shows the corresponding mass distribution of entropy generated ($\Delta S$) per $\log(M)$. We color-code the regions for PISN mass gap (yellow; $45\text{--}130~M_\odot$), while the remaining mass range (light shading) denotes the stellar-collapse regime outside the gap. Annotations report the integrated $\Sigma~\Delta S$ contributed by the two mass regimes.}
\label{fig:mass}
\end{figure}

Under this assumption, the BBH number density distribution over $5-300~M_\odot$ (blue curve in Fig. \ref{fig:mass}) closely follows the PP model calibrated to the GWTC-4 catalog. The corresponding entropy distribution (red curve in Fig. \ref{fig:mass}) inherits the characteristic double-peaked structure. In both cases the lower-mass mode dominates, but the peak-to-peak asymmetry is substantially more pronounced for the number-density distribution than for the entropy-production distribution. Specifically, at $M = 33.6~M_\odot$, the entropy increase peaks at a $\Delta S = 1.07 \times 10^{90}~k$, whereas at $M = 9.84~M_\odot$, the entropy increase peaks at $\Delta S = 4.44 \times 10^{90}~k$. As described by Eq. \ref{eq:entropy}, the entropy of a black hole scales with $S \propto M^2$. Consequently, the higher mass peak near $35~M_\odot$ is intrinsically more entropy-productive per event, but this is partially offset by its relatively lower number density compared to the peak at the lower mass near $10~M_\odot$. This interplay between mass weighting and event abundance shapes the overall entropy distribution of the observed population.

BBHs within the PISN mass gap contribute $\sim 25\%$ of the total entropy increase, even though they comprise only $1.3\%$ of the total BBH sample. This disparity highlights the outsized impact of rare, high-mass (and therefore high-entropy) events on the total merger-generated entropy budget.

\begin{figure}[t!]
\centering
\includegraphics[width=0.5\textwidth]{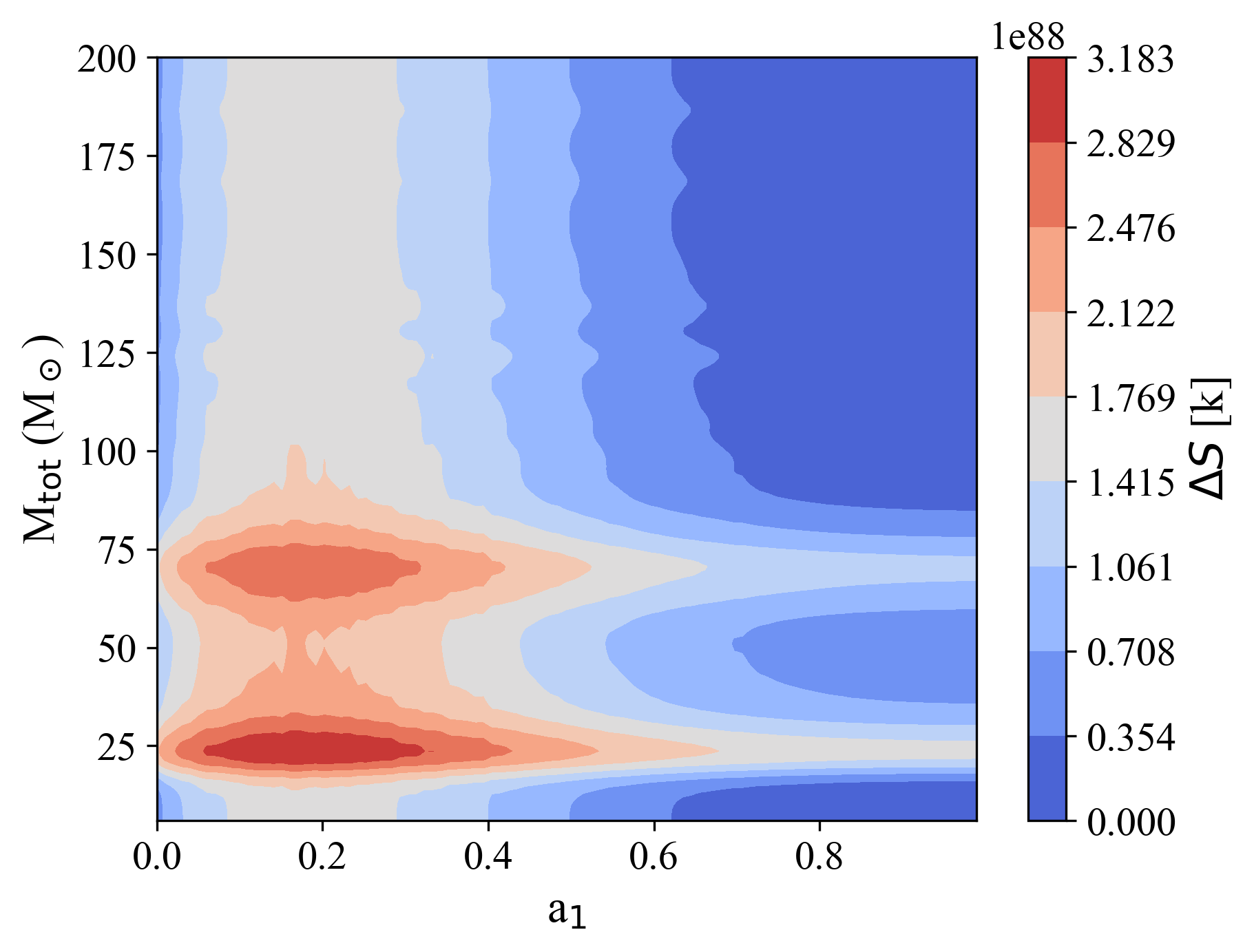}
\caption{\textbf{Entropy Production in Mass-Spin Space}: the total entropy change $\Delta S$ from stellar BBH mergers across all redshift $z \in [0,20]$ in the mass ($M_\mathrm{tot}$) - spin ($a_1$) parameter space.} 
\label{fig:spin}
\end{figure}

In  Fig. \ref{fig:spin}, we show the population-weighted total entropy increase $\Delta S$ mapped onto the $(M_\mathrm{tot}, a_1)$ parameter space and integrated over redshift $z \in [0,20]$. The resulting structure reflects the underlying mass and spin distributions inferred from the GWTC-4 population. Because we work in total mass, we marginalize over the population distribution of the mass ratio $q$, which favors nearly equal-mass binaries ($q \simeq 1$). With this $q$-weighing, the $\Delta S$ contribution associated with the higher-mass peak ($\sim 35~M_\odot$) becomes less dominant than the primary mass representation shown in Fig. \ref{fig:mass}. While entropy production remains similarly significant across the mass range $20 - 80~M_\odot$, the double-peak structure remains obvious. 

In comparison, the spin magnitude has a weaker influence on $\Delta S$ than the masses. As indicated by Eq. \ref{eq:entropy}, the dominant scaling is $\propto M^2$, while the spin dependence is subdominant. Although the Bekenstein–Hawking entropy of an individual black hole is maximized for zero spin, the rate-weighted $\Delta S$ peaks around $a \sim 0.2$, consistent with the LVK-inferred spin distribution. This indicates that spin modulates the merger entropy budget, but the overall rate-weighted entropy production is driven primarily by the mass distribution.

\begin{figure}[t!]
\centering
\includegraphics[width=0.48\textwidth]{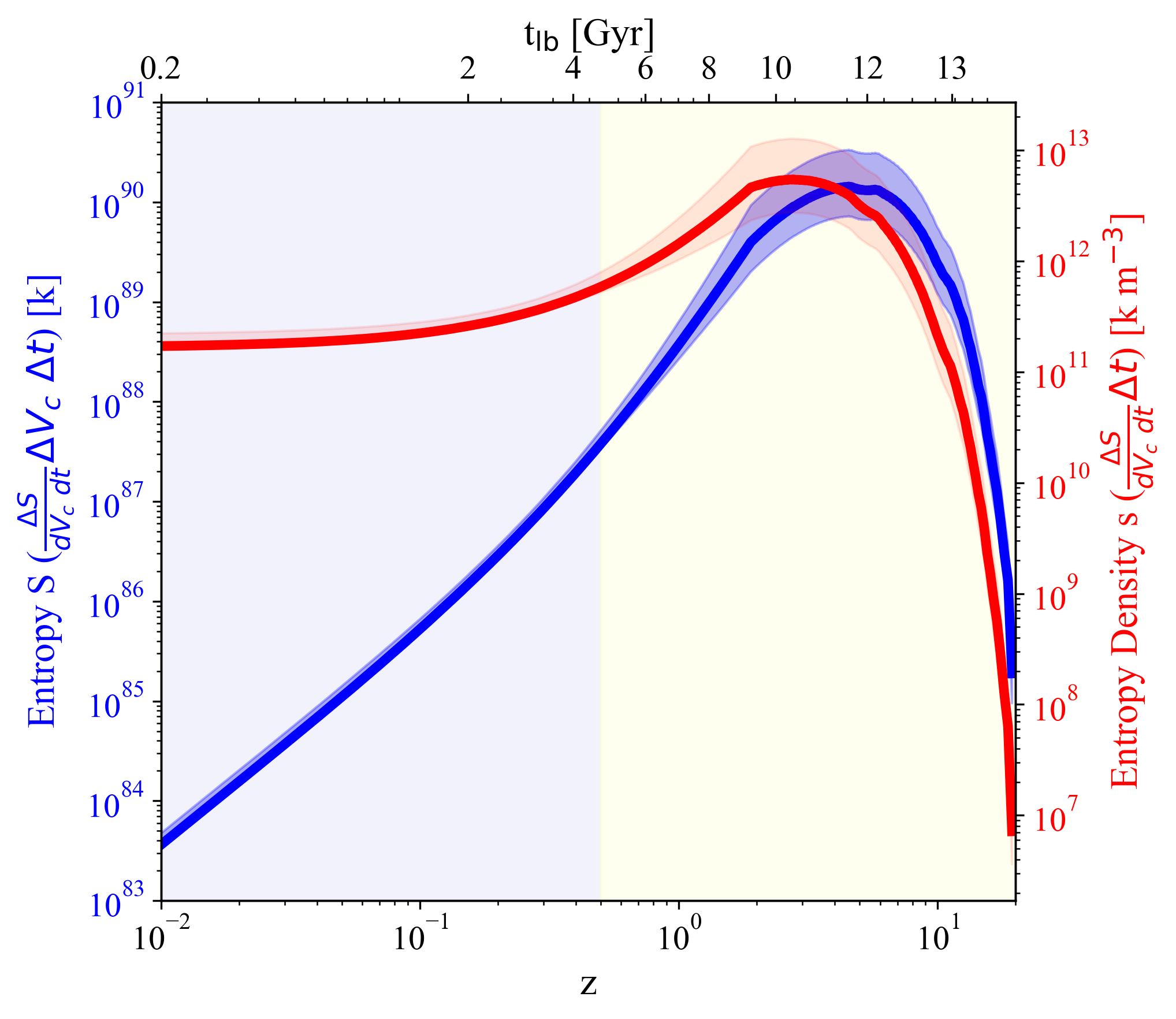}
\caption{\textbf{Cosmic History of Merger-Generated Entropy}: total entropy (left y axis \& blue curve) and entropy density (right y axis \& red curve) from merging BBHs in the comoving frame across $z \in [0.01, 20]$, as a function of the redshift $z$ (bottom x axis) and of the look-back time $t_{lb}$ (top x axis). 
The dark energy-dominated and matter-dominated eras are shaded purple and yellow, respectively.}
\label{fig:growth}
\end{figure}

Finally, we infer the redshift evolution of the merger-generated entropy using Eq. \ref{eq:growth}. The resulting total contribution per bin is shown by the blue curve in Fig. \ref{fig:growth}. We also define a localized entropy density $s(z)$ by excluding the differential comoving volume factor $\Delta V_c$, so that $s(z)$ isolates the intrinsic redshift dependence of the merger rate. $s(z)$ scales approximately with the merger rate scaled by a mass-weighted entropy-per-event factor given our uniform $t_\mathrm{lb}$ binning. Consequently, the red curve closely follows the BBH merger rate history and peaks at $z \simeq 2.79$. By contrast, the total entropy $S(z)$ retains the comoving-volume weighting through $\Delta V_c$ and therefore represents the full contribution from each redshift shell. In this sense, $S(z)$ effectively traces a volume-weighted rate, approximately $\mathcal{R}(z) \Delta V_c$. As a result, the maximum of the blue curve is shifted to higher redshift, with $S(z)$ peaking at $z\simeq 4.55$. Physically, this peak offset reflects a competition between the decline of the merger rate at high redshift and the rapid growth of the available comoving volume per redshift interval: the total contribution continues to rise beyond the intrinsic merger-rate maximum until the rate suppression outweighs the volume growth.

To quantify the redshift scaling in different regimes, we approximate the growth curve with piecewise power laws in redshift over the monotonic segments of the evolution. In the low-redshift (dark energy-influenced) regime $z \lesssim 0.5$, the total entropy growth is well described by a power law $S_\mathrm{growth} = 1.46 \times 10^{88} \times z^{2.37}$ with a tightly constrained slope from the GWTC-4 population uncertainty. At intermediate redshift ($0.5 \lesssim z \lesssim4.5$), the growth steepens ($S \propto z^{3.07}$), culminating in a maximum value of $S_\mathrm{peak} = 1.43^{+1.91}_{-0.71} \times 10^{90}~k$ at $z\simeq4.55$. Beyond the peak, the total entropy drops rapidly ($S \propto z^{-4.82}$), consistent with the sharp suppression of the merger rate at higher redshift.

Building on our new calculations, we revisit the entropy census of major cosmic components using the budget compiled by \cite{Egan_2010} as a reference baseline. Over the past decade, surveys and missions have substantially refined several inputs to this census, including the observational confirmation by the LVK Collaboration of heavier stellar black holes in the $\sim 45-130~M_\odot$ range. Incorporating these updates, together with additional components considered in more recent work, we present a revised and expanded entropy budget of the observable universe in Table \ref{tab:budget}. This compilation extends earlier inventories \citep{Frautschi1982, Frampton_2008, frampton2009entropyintermediatemassblackholes, Frampton_2009, Egan_2010, profumo2024newcensusuniversesentropy} by adopting updated estimates where available and providing references for each entry. In particular, we explicitly include the contribution associated with BBH mergers, which emerges as a substantial component of the late-time thermodynamic budget.

\begin{figure*}[t!]
\centering
\includegraphics[width=0.9\textwidth]{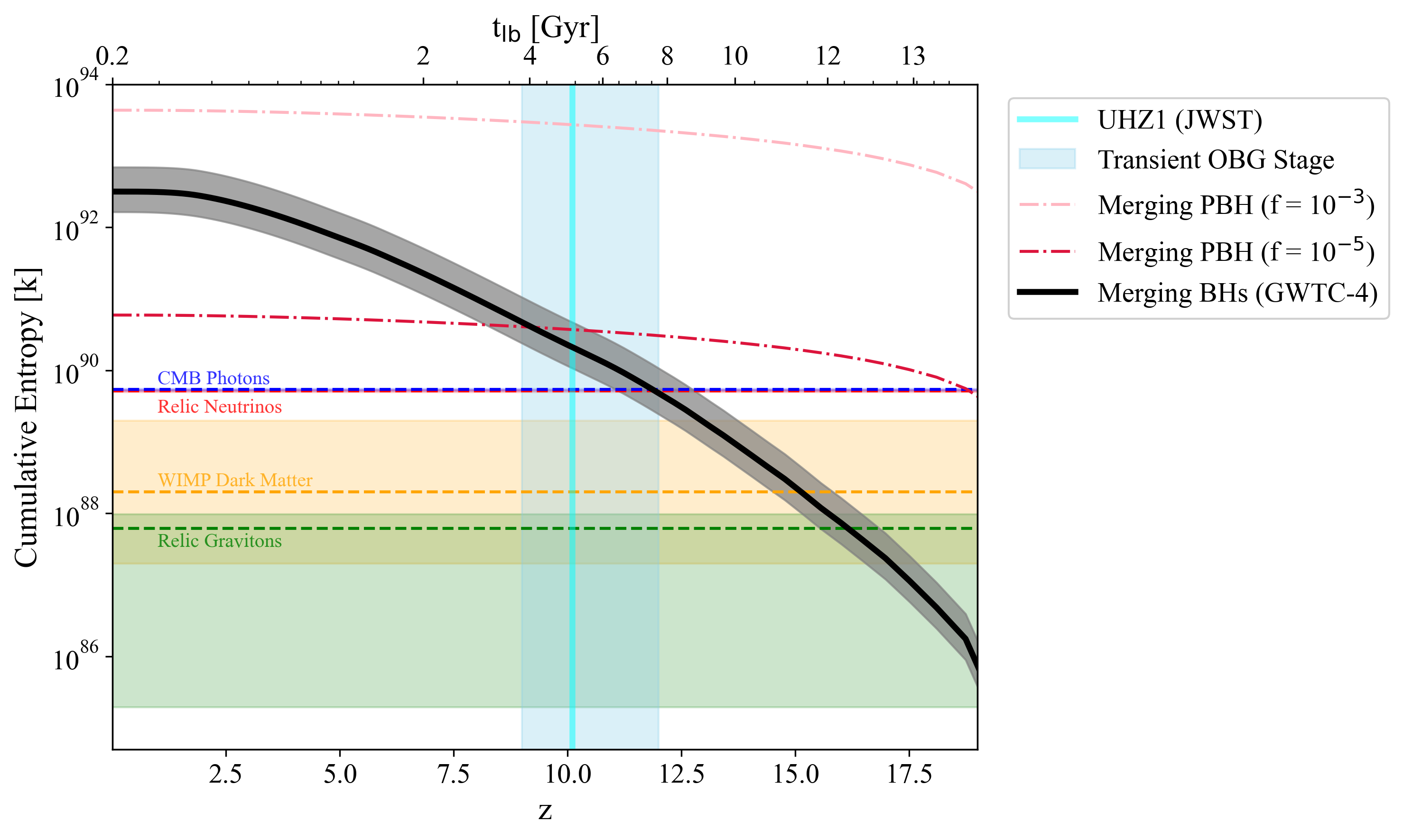}
\caption{\textbf{Cumulative Cosmological Entropy Budget}: cumulative entropy from merging BBHs in the comoving frame as a function of redshift over $z \in [0.01,20]$. For comparison, we also show the entropy contribution from PBHs for two PBH fraction choices. For reference, the entropy contributions from other cosmic components, as estimated in \cite{Egan_2010}, are included. The vertical line marks the redshift of the most distant black hole observed by JWST in galaxy UHZ1, and the shaded vertical band indicates the transient OBG stage associated with this epoch. \label{fig:pbh?}}
\end{figure*}


\section{Cosmological Implications}
\label{sec:cumulation+rate}

To evaluate the cumulative entropy contribution from BBH mergers since the Big Bang, we integrate the differential entropy growth (blue curve in Fig. \ref{fig:growth}) from the highest redshifts down to the present epoch. The resulting cumulative entropy $\mathcal{S}_{\rm BH}(z)$ is shown in Fig. \ref{fig:pbh?}. This metric captures the cumulative, irreversible entropy from BBH mergers across cosmic time, quantifying their contribution to the cosmic entropy budget from the earliest structure formation epochs to today.

\subsection{The Thermodynamic Crossover at Cosmic Dawn}
\label{subsec:cmb_cross}
As shown in Fig. \ref{fig:pbh?}, the cumulative entropy increases monotonically, as expected for an integrated entropy-generation history. A salient feature of this evolution is the ``thermodynamic crossover,'' where the cumulative entropy produced by mergers exceeds the total entropy of the CMB photons. We find that this transition occurs at a redshift of $z = 12.6^{+1.5}_{-3.5}$. This result indicates that the cumulative merger-associated entropy becomes comparable to, and then exceeds, the CMB photon entropy budget significantly earlier than the peak of cosmic star formation, suggesting that merger-driven entropy accumulation is already substantial during Cosmic Dawn in this reference comparison.

This crossover is particularly intriguing in the context of high-redshift observations. The James Webb Space Telescope (JWST) has reported the detection of an accreting black hole with mass $M_\mathrm{BH} \sim 4 \times 10^7~M_\odot$ in the galaxy UHZ1 at $z \sim 10.1$ \citep{natarajan2023detectionovermassiveblackhole}. Theoretical models predict a transient phase of over-massive black hole galaxies (OBGs) over $z \in [9,12]$ \citep{2013MNRAS.432.3438A, Natarajan_2017}, which is highlighted in the figure (blue shaded region). Intriguingly, the cumulative entropy from BBH mergers is found to exceed that of the CMB around the onset of this transient OBG stage. While this temporal proximity does not imply a causal connection, it is consistent with scenarios in which BBH mergers contribute a non-negligible share of the cumulative entropy generated by $z \gtrsim 10$. In particular, it may be compatible with light-seed models in which Population III remnants produce black hole seeds in the mass range $10$-$100~M_\odot$. These seeds form at very high redshift, participate in hierarchical mergers, and may contribute to the assembly of the massive black holes inferred in JWST observations. The strength of this inference, however, depends on the assumed high-redshift BBH merger rate and on the entropy-generation prescription adopted for mergers.

\subsection{Constraints from the Dark Ages: The PBH Floor}
Our analysis primarily focuses on the stellar-origin BBH population, which is expected to be negligible at very high redshift, prior to the onset of Population III star formation (and our results are shown over the range highlighted in the data $z\le20$). In the absence of additional channels, the merger-generated entropy in this regime therefore approaches a minimal baseline set by the stellar contribution alone. If PBHs constitute a non-zero fraction of the dark matter, PBH binaries can form and merge well before the first stars ignite, thereby establishing an ``entropy floor” for the cumulative merger-generated gravitational entropy at high redshift.

The magnitude of this floor depends sensitively on the PBH abundance, $f_\mathrm{PBH} \equiv \Omega_\mathrm{PBH}/\Omega_\mathrm{DM}$, and on the PBH mass function. Low-mass PBHs ($M \ll 1~M_\odot$) contribute negligibly to the merger-generated entropy budget considered here even for $f_\mathrm{PBH} \sim 1$ because the characteristic black-hole entropy scales steeply with mass. By contrast, stellar-mass PBHs in the LVK-relevant range ($M \sim 30~M_\odot$) carry a much larger entropy per merger event and can substantially enhance the cumulative entropy from mergers. In this study, we consider two representative abundances: a higher-abundance case ($f_\mathrm{PBH} = 10^{-3}$, pink dot-dashed line) and a conservative low-abundance case ($f_\mathrm{PBH} = 10^{-5}$, red dot-dashed line). The resulting cumulative entropy histories are shown in Fig.~\ref{fig:pbh?} and compared directly to the stellar-origin BBH contribution.

As illustrated by the $f_\mathrm{PBH} = 10^{-3}$ case, the PBH merger contribution exceeds the stellar-origin BBH contribution across the full $z$ range shown, implying that the merger-generated entropy budget is PBH-dominated in this scenario. This behavior is a direct consequence of the efficient early-time mergers of primordial binaries, with a rate scaling $\mathcal{R}(t) \propto t^{-34/37}$ \citep{Sasaki_2016}, which enables substantial entropy accumulation already at very high redshift. Consequently, the stellar channel remains sub-dominant in cumulative merger-generated entropy at all epochs displayed.

The lower-abundance case ($f_\mathrm{PBH} = 10^{-5}$) yields a more dynamic thermodynamic history. Even at this level, PBH mergers can accumulate entropy early, surpassing the entropy budget of CMB photons as early as $z \sim 18.7$. As cosmic structure formation proceeds, the stellar-origin BBH merger rate rises rapidly, and the cumulative stellar contribution eventually intersects and overtakes the PBH contribution. We find a crossover at $z=9.17^{+1.31}_{-1.01}$, after which the cumulative merger-generated entropy is driven primarily by astrophysical black holes. This delineates two regimes in the merger-generated entropy history for $f_\mathrm{PBH} \sim 10^{-5}$: a PBH-dominated phase at earlier times followed by a stellar-dominated phase toward lower redshift.

\subsection{Cosmological Density Parameters \& Thermodynamic Asymmetry}
Beyond the entropy census, we compute the cosmological density parameters of the black hole population to quantify the energetic footprint.

We first estimate the mass-energy inventory stored in stellar-origin black holes. We find an initial mass density parameter of $\Omega_\mathrm{BH} = 5.32^{+6.65}_{-3.37}\times 10^{-5}$, which is consistent with typical local-universe estimates ($\sim 10^{-5}$). The modest offset is expected given our assumption of a uniformly low-metallicity population ($Z=2\times 10^{-4}$): weaker line-driven winds reduce stellar mass loss, leading to systematically heavier compact remnants and a larger inferred black hole mass density.

Only a small fraction of this population undergoes mergers. We infer the mass density in merger remnants to be $\Omega_\mathrm{BH,~rem} = 4.19^{+5.68}_{-2.44} \times 10^{-9}$. The ratio $\Omega_\mathrm{BH,~rem}~/~\Omega_\mathrm{BH} \sim 8 \times 10^{-5}$ indicates that only a tiny fraction of the stellar black-hole mass budget has been processed through BBH mergers to date. Integrating the merger history, we find a GW energy density parameter of $\Omega_\mathrm{GW} = 5.01^{+6.50}_{-2.85}\times 10^{-11}$, approximately six orders of magnitude below the energy densities of standard relativistic components, such as CMB photons ($\Omega_{\gamma} \approx 5 \times 10^{-5}$) or neutrinos ($\Omega_{\nu} \approx 3.4 \times 10^{-5}$).

Thus, although black holes dominate the late-time entropy budget, the Universe’s energy budget remains overwhelmingly carried by radiation rather than by merger-generated GWs. The origin of this disparity is that BBH mergers radiate only a small fraction of the system’s mass-energy, with most of the energy remaining behind the horizon in the remnant black hole. We find $\Omega_\mathrm{GW}/\Omega_\mathrm{BH,rem} \sim 10^{-2}$, consistent with the modest radiative efficiency of BBH mergers. Yet even this small radiated fraction can accompany a large entropy increase in accord with the second law. These results illustrate a \textit{thermodynamic asymmetry} in black hole mergers:
\begin{enumerate}
    \item \textbf{Energetically Inefficient:} BBH mergers radiate only a minute fraction of their mass-energy ($\Omega_\mathrm{GW}/\Omega_\mathrm{BH,rem} \sim 10^{-2}$) and contribute negligibly to the universe's energy budget compared to radiation or matter.
    \item \textbf{Entropically Dominant:} Conversely, as demonstrated in Sec. \ref{subsec:cmb_cross}, these same events drive a disproportionate increase in cosmic entropy, surpassing the CMB photon entropy budget at $z \sim 12$.
\end{enumerate}
Therefore, we interpret $\Omega_\mathrm{GW}$ not as a major energy component, but as the unavoidable energetic by-product of an irreversible process that efficiently advances the universe toward a higher-entropy state.

\subsection{Volume-normalized Retrospective Entropy Density}

To characterize the volumetric accumulation of entropy, we compute the cumulative number of BBH mergers enclosed within $z$ by integrating the merger-rate density over the past light cone:

\begin{equation}
    \centering
    N_\mathrm{BBH}(z) = \int_{0}^{z} \mathcal{R}(z)~dV_c(z)~dt_\mathrm{lb}(z)
    \label{eq:cumulative}
\end{equation}

We then introduce a new quantity termed \textit{retrospective entropy density}, a volume-normalized cumulative measure that compares the total merger-generated entropy integrated over $z \in [0.01, z_0]$ to the comoving volume enclosed within the same interval:

\begin{equation}
    \mathrm{Retrospective~Entropy~Density~(z_0)} = \frac{\sum^{\mathrm{N_{BBH}(z_0)}}_{i=1} \Delta S_i}{\int^{z_0}_0 dV_{c}(z)}
    \label{eq:rate_entropy_density}
\end{equation}

\begin{figure}[t!]
\includegraphics[width=0.49\textwidth]{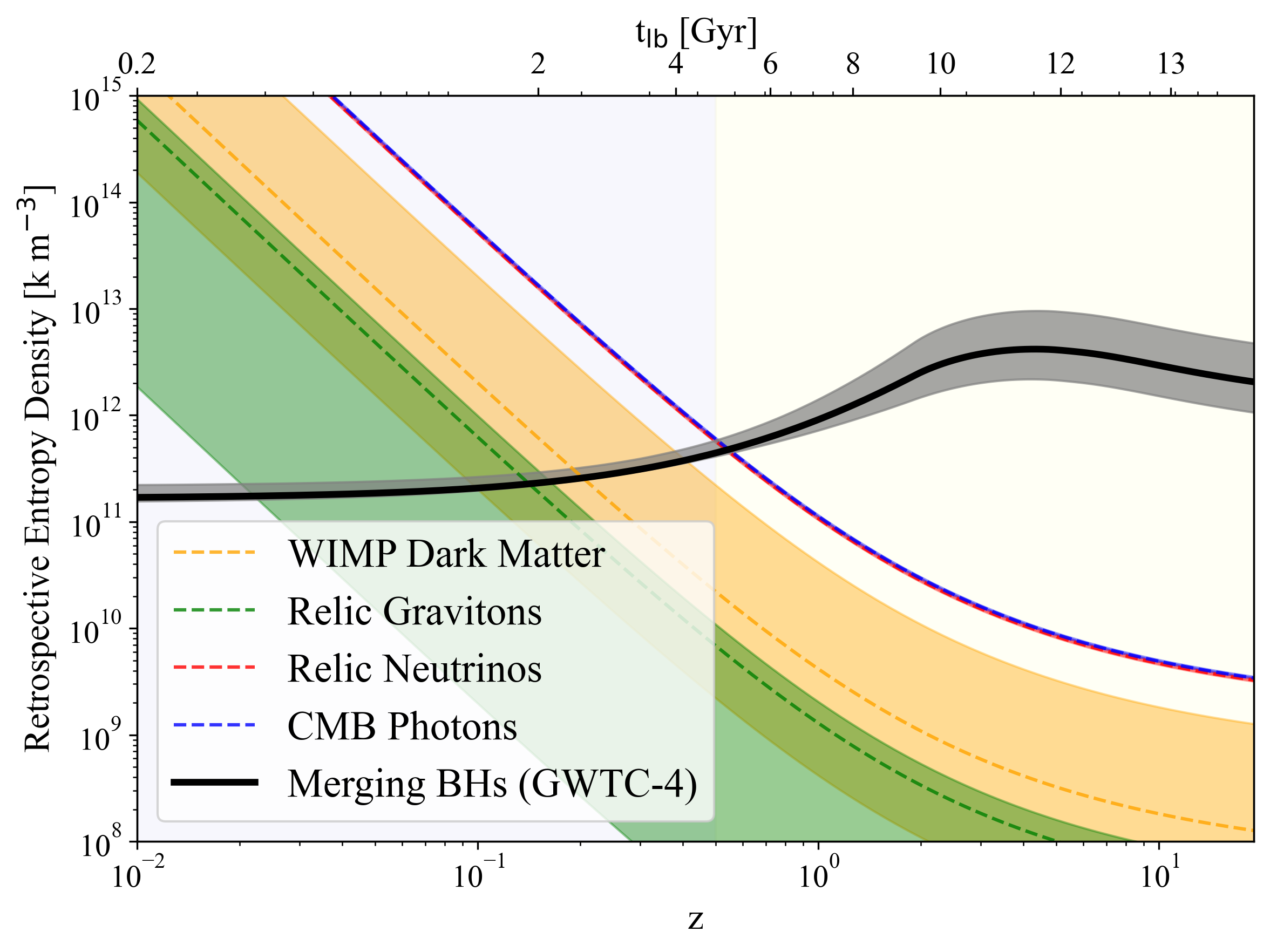}
\caption{\textbf{Volume-normalized Retrospective Entropy Density}: retrospective entropy density as a function of redshift $z$ (bottom x axis) and look-back time $t_{lb}$ (top x axis), constructed by normalizing the cumulative entropy budget integrated over $z\in [0.01,20]$ by the comoving volume enclosed within the same interval. For reference, we include the entropy contributions of several conserved cosmic components reported in \cite{Egan_2010}. The purple-shaded region corresponds to the dark energy–dominated era ($z \le 0.5$), while the yellow-shaded region represents the matter-dominated era ($z > 0.5$).
\label{fig:rate}
}
\end{figure}

Geometrically, this definition corresponds to constructing concentric comoving spheres centered on the observer (the present epoch), each extending out to a maximum redshift $z_0$. The numerator represents the cumulative entropy produced by all BBH mergers occurring within that past-light-cone volume, while the denominator provides the corresponding comoving volume normalization. We emphasize that this quantity is \emph{not} the instantaneous physical entropy density at redshift $z_0$. Rather, it serves as a retrospective accounting metric, quantifying the average density of merger-driven entropy associated with the spacetime region enclosed by $z_0$. The resulting evolution is shown in Fig. \ref{fig:rate}.

Since both the cumulative entropy $\Sigma~\Delta S$ and the enclosed comoving volume $V_c(<z)$ increase monotonically with redshfits, their ratio elucidates the competition between entropy accumulation and the expansion of the accessible volume. In our reconstruction based on GWTC-4, the retrospective entropy density rises from the local Universe to reach a maximum of $4.17^{+5.35}_{-2.00} \times 10^{12}~k~m^{-3}$ at $z = 4.33^{+0.08}_{-0.12}$, before declining toward higher redshifts. This peak reflects the interplay between (i) the growing cumulative contribution from mergers as the integration extends to earlier cosmic times, and (ii) the rapid geometric growth of the enclosed comoving volume, which dilutes the volume-normalized ratio once the merger contribution ceases to grow at a comparable rate. As $z$ approaches the upper bound ($z \simeq 20$), the retrospective entropy density asymptotes to the average comoving entropy density contributed by BBH mergers in the observable universe, consistent with the value reported in Table \ref{tab:budget}.

\label{subsec:retro_den}

For context, Fig. \ref{fig:rate} shows the profiles of several conserved cosmic components using entropy budgets compiled in \cite{Egan_2010}. For these relic populations, the total entropy budget is approximately conserved following Big Bang Nucleosynthesis, so their retrospective entropy density under Eq. \ref{eq:rate_entropy_density} is obtained by distributing a fixed total entropy over an increasing enclosed comoving volume $V_c(<z_0)$. Consequently, their curves decrease monotonically with increasing $z_0$, with the evolution driven entirely by the comoving volume normalization rather than intrinsic entropy production.

Under this retrospective normalization, the curve for BBH-merger becomes comparable to and intersects the baselines for relic photons and relic neutrinos near $z=0.55^{+0.01}_{-0.05}$, broadly coinciding with the transition between dark energy–dominated and matter-dominated eras. We stress that this intersection should be interpreted strictly within the context of the adopted volume-normalized cumulative definition. It signifies that, when averaged over the local comoving volume enclosed within $z \approx 0.5$, the integrated entropy produced by astrophysical black hole mergers is sufficient to rival the volume-diluted density of relic radiation backgrounds.

\section{Conclusion}
In this study, we have presented a revised cosmic entropy budget, explicitly quantifying the contributions from stellar-collapse black holes and their mergers. By integrating LVK event-level posteriors with GWTC-3/4 population inference and cosmological merger rate histories, we constructed an inventory driven primarily by the mass spectrum—a direct consequence of the $S \propto M^2$ scaling inherent to the Hawking area theorem. Utilizing the GWTC-4 parameter space, we characterized the dependence of entropy production on mass, spin, and redshift, revealing three key insights. First, we identified a critical "thermodynamic crossover" during Cosmic Dawn ($z \sim 12$). This epoch, coinciding with the predicted OBG phase, marks the moment when the cumulative entropy generated by BBH mergers surpasses the entropic content of the CMB photon background. Second, we demonstrated that even a trace population of primordial black holes would establish an early "entropy floor," dominating the thermodynamic landscape of the Dark Ages. Finally, by computing the cosmological density parameters, we highlighted a profound thermodynamic asymmetry: while BBH mergers are energetically inefficient, radiating only a minute fraction of their mass-energy as gravitational waves, they drive a disproportionate irreversible increase in cosmic entropy.

We acknowledge that our high-redshift conclusions rely on extrapolating the BBH merger rate from \cite{Mapelli_2018} beyond the horizon of current ground-based detectors, leaving $\mathcal{R}(z)$ less constrained at $z\gtrsim2$. Future GW facilities will provide the necessary empirical anchor to test these predictions. Space-based observatories like LISA~\cite{LISAstudyreport} will access lower-frequency signals from massive and IMBH systems out to high redshifts \citep{Rosado_2016, Ng_2022, Mancarella_2023, Wu_2024}, while proposed mid-band concepts such as the Laser Interferometer Lunar Antenna (LILA) would bridge the gap between LVK and LISA, extending sensitivity to heavier binaries and earlier cosmic times \citep{jani_2021, jani2025laserinterferometerlunarantenna, LILAPioneer, LILAHorizon}. Together, these missions will directly map $\mathcal{R}(z)$ over a broader mass range, reducing reliance on star formation rate-based extrapolations and enabling a precise reconstruction of entropy production in the Universe's earliest epochs.

\section{Acknowledgement}
We thank Priyamvada Natarajan for insightful discussion. S.C. acknowledges the support from the Littlejohn Fellowship, Vanderbilt Immersion Program and the Vanderbilt University Summer Research program. K. J.’s work was supported in part by the Lunar Labs Initiative at Vanderbilt
University, which is funded by grants from the John Templeton Foundation and the Vanderbilt Scaling Grant from the Office of the Vice Provost for Research and Innovation. This material is based upon work supported by NSF’s LIGO Laboratory, which is a major facility fully funded by the National Science Foundation. This research has made use of data, software and / or web tools obtained from the Gravitational Wave Open Science Center (\url{https://www.gw-openscience.org/}), a service of LIGO Laboratory, the LIGO Scientific Collaboration and the Virgo Collaboration. LIGO Laboratory and Advanced LIGO are funded by the United States National Science Foundation (NSF) as well as the Science and Technology Facilities Council (STFC) of the United Kingdom, the Max-Planck-Society (MPS), and the State of Niedersachsen/Germany for support of the construction of Advanced LIGO and construction and operation of the GEO600 detector. Additional support for Advanced LIGO was provided by the Australian Research Council. Virgo is funded, through the European Gravitational Observatory (EGO), the French Centre National de Recherche Scientifique (CNRS), the Italian Istituto Nazionale di Fisica Nucleare (INFN), and the Dutch Nikhef, with contributions by institutions from Belgium, Germany, Greece, Hungary, Ireland, Japan, Monaco, Poland, Portugal, Spain.

\bibliography{sample631}

\end{document}